\documentclass[floatfix,aps,prl,twocolumn,nofootinbib,superscriptaddress]{revtex4}

\usepackage{graphicx}
\usepackage{subfigure}
\usepackage{epsfig}
\usepackage{amsmath}
\usepackage{amsfonts}
\usepackage{amssymb}
\usepackage{hyperref}
\usepackage{dsfont}
\usepackage{color}
\usepackage{soul}
\usepackage[english]{babel}

\newcommand{\ket}[1]{|{#1}\rangle}
\newcommand{\bra}[1]{\langle{#1}|}

\newcommand{\cala}{\mathcal A}
\newcommand{\cals}{\mathcal S}
\newcommand{\calu}{\mathcal U}

\usepackage[normalem]{ulem}

\setstcolor{red}

\addto\captionsenglish{}

\newcommand{\braket}[2]{\langle{#1}|{#2}\rangle}
\newcommand{\beq}{\begin{equation}}
\newcommand{\eeq}{\end{equation}}
\newcommand{\beqa}{\begin{eqnarray}}
\newcommand{\eeqa}{\end{eqnarray}}

\newcommand{\tr}{\mathop{\mathrm{Tr}}\nolimits}

\def\id{\mathbb{I}}

\newcommand{\uba}{Departamento de F\'\i sica, FCEyN, UBA, Pabell\'on 1,
Ciudad Universitaria, 1428 Buenos Aires, Argentina}
\newcommand{\ifiba}{Instituto de F\'\i sica de Buenos Aires, UBA CONICET,
Pabell\'on 1, Ciudad Universitaria, 1428 Buenos Aires, Argentina}
\newcommand{\bgu}{Department of Physics, Ben-Gurion University of the Negev, Be'er Sheva 84105, Israel.}
\newcommand{\amsterdam}{Van der Waals-Zeeman Institute, University of Amsterdam, Science Park 904, PO Box 94485, 1090 GL Amsterdam, The Netherlands}

\begin{document}

\title{Using a quantum work meter to
test non-equilibrium fluctuation theorems}

\author{Federico Cerisola} \email[Correspondence and requests for materials should be addressed to F.C. at: ]{cerisola@df.uba.ar} \affiliation{\uba} \affiliation{\ifiba}
\author{Yair Margalit} \affiliation{\bgu}
\author{Shimon Machluf} \affiliation{\amsterdam}
\author{Augusto J. Roncaglia} \affiliation{\uba} \affiliation{\ifiba}
\author{Juan Pablo Paz} \affiliation{\uba} \affiliation{\ifiba}
 \email[]{paz@df.uba.ar} 
\author{Ron Folman} \affiliation{\bgu}

\date{\today}

\begin{abstract}

Work is an essential concept
in classical thermodynamics, and in the quantum regime,
where the notion of a trajectory is not available,
its definition is not trivial. For driven (but otherwise isolated) 
quantum systems, work can be defined
as a random variable, associated with the 
change in the internal energy. The probability
for the different values of work captures
essential information describing the behaviour of the
system, both in and out of thermal equilibrium. 
In fact, the work probability distribution is at the core of ``fluctuation theorems'' in quantum thermodynamics. 
Here we present the design and implementation of a quantum work 
meter operating on an ensemble of cold atoms, which are controlled by an atom chip. Our device not
only directly measures work but also directly
samples its probability distribution. 
We demonstrate the operation of this 
new tool and use it to verify the validity of 
the quantum Jarzynksi identity.
\end{abstract}

\maketitle

\section{Introduction}
Classical fluctuation theorems establish surprising relations between non-equilibrium and
equilibrium concepts. In particular, the work performed on a system during non-equilibrium processes is connected with key concepts of
equilibrium thermodynamics, such as the free-energy \cite{jarzynski1997nonequilibrium,crooks1999entropy}.
These relations have been verified in various experiments involving microscopic  
thermodynamic systems  \cite{hummer2001free,liphardt2002equilibrium,collin2005verification}.
Recent advances in quantum technologies enable the control of small quantum systems that can be 
manipulated far from the regime where  the usual thermodynamical laws are obeyed. This triggered
the development of the rapidly growing field of non-equilibrium quantum thermodynamics  \cite{esposito2009nonequilibrium,campisi2011colloquium, seifert2012stochastic,goold2016role}. 

When quantum fluctuations dominate, defining and measuring work
and heat, two central concepts in classical thermodynamics, is
non-trivial. For driven, but otherwise isolated, quantum systems, work $w$ is a
random variable associated with the change in the internal energy \cite{talkner2007fluctuation}, as the first law of thermodynamics indicates.
Thus, the commonly accepted definition of quantum work requires a two-time measurement strategy,
which consists of performing two projective energy
measurements, one at the beginning and the other at the
end of the process.
Then, work is associated with the measured energy difference.
However, implementing the two-time measurement 
is experimentally difficult \cite{huber2008employing,watanabe2014generalized}
due to the fact that the two projective measurements are unavoidably disruptive
(see Ref.  \cite{an2015experimental} for an ion trap implementation). 
Alternative methods to evaluate the work probability
distribution that rely on the direct estimation of its Fourier transform were
also proposed in Refs.  \cite{mazzola2013measuring,dorner2013extracting} and
later implemented in NMR experiments \cite{batalhao2014experimental}.

In this paper we present the design and the  experimental implementation of a ``quantum work meter'' (QWM)
operating on an ensemble of cold atoms, combining the idea presented in Ref. \cite{roncaglia2014work} and the
experimental setup used in Ref. \cite{machluf2013coherent}. 
Our QWM is conceptually different from previous work-measurement devices. 
Its main advantage is that the QWM efficiently samples $P(w)$, which is a direct observable in the experiment.
Namely, our QWM not only directly measures work but also directly samples its probability
distribution $P(w)$ [i.e. the outcome $w$ is obtained with probability $P(w)$].  As the work probability distribution
plays a central role in the fluctuation theorems of non-equilibrium quantum thermodynamics,
the QWM is an ideal tool to  test their validity. In particular,
we use it to verify the Jarzynski  identity \cite{jarzynski1997nonequilibrium,tasaki2000jarzynski, kurchan2000quantum, mukamel2003quantum, talkner2007fluctuation}.\\

\begin{figure*}[ht]
    \centering
    \includegraphics[width=0.8\linewidth]{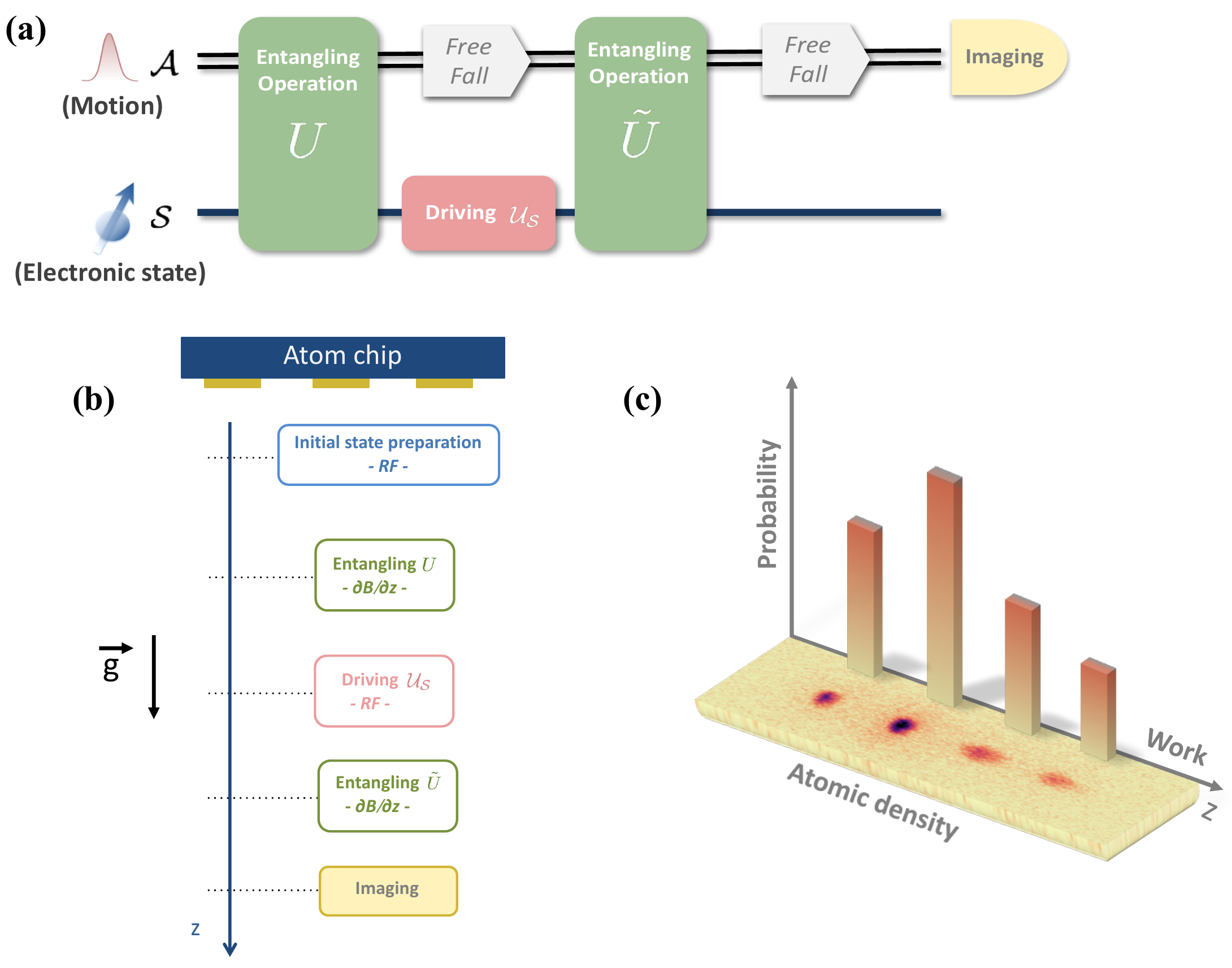}
    \caption{\label{fig:esq-povm-meas} The Quantum Work Meter.
    {\bf (a})~A quantum circuit for the Quantum Work Meter (QWM). $\cals$ and $\cala$ are entangled so that the eigenvalue
of the observable $H$ of the system
$\cals$ is coherently recorded by $\cala$. Then $\cals$ is driven by $\calu_\cals$.
Finally, another entangling operation between $\cals$ and $\cala$ creates a record of $w$ on
$\cala$. In the experiment, $\cala$ is encoded in the motional degree of freedom of
the atoms along the vertical direction $z$, which also evolves while freely falling. $\cals$ is the
pseudospin associated with  the Zeeman sub-levels of a
$^{87}$Rb atom.
        {\bf (b)}~Physical operations for the QWM on
an atom chip:
        i)~The atoms, prepared  in state  \(\ket{2}\), are released from the trap,
and a RF field  generates an initial pseudo-thermal state.
        ii)~After $2.4$\,ms, internal and motional degrees of freedom are entangled with a magnetic gradient pulse ($U$), applied 
        for a duration of $\tau= 40\,\mu$s.
        iii)~Another RF field drives $\cals$.
        iv)~3.1~ms after the application of $U$, a second magnetic gradient pulse ($\tilde U$) is applied for a duration of $\tilde \tau= 300\,\mu$s. At this stage, $\cala$ keeps a record of
the different work values.
        v)~ After $18.2\,$ms from the application of $\tilde U$,
        the positions and optical densities of the atomic clouds are measured. The number of atoms in each cloud  reveals the
        work probability in a single experimental realisation.
        {\bf (c)}~Image of the four clouds obtained at the end of
a single run of the QWM. The four possible values of $w$ fix the position of each cloud. }
\end{figure*}

\section{Results}

\noindent\textbf{Work measurement and the QWM.}
A QWM is an apparatus that  measures the work performed on a driven quantum system
whose Hamiltonian varies from an initial $H$ to a final $\tilde H$ with  eigenvalues  $E_n$ and $\tilde E_m$, respectively.
For an isolated system $\cals$, with a $D$-dimensional space of states, the number of different
values of work is bounded by $D^2$. Therefore, the pointer of the QWM has $D^2$ distinct positions (one for each value of
$w=w_{nm}=\tilde E_m-E_n$).  The QWM presented here
enables  us to choose $H$ and $\tilde H$
(fixing the possible values of $w$) and to
vary the intermediate driving (inducing
different evolution operators denoted
as $\calu_\cals$). In this way, we vary the
probability $P(w)$, which depends on the
intermediate driving $\calu_\cals$.

By sampling $P(w)$, we use the QWM  to verify
a fundamental result in non-equilibrium quantum thermodynamics: the Jarzynski identity. 
This identity states that for any initial state with populations identical to the ones associated to 
a thermal Gibbs state and for any distribution $P(w)$, the
linear combination ${\langle e^{-\beta w}\rangle = \sum_w e^{-\beta w} P(w)}$,
where $\beta=1/k_BT$ is the inverse temperature of the system, 
is an equilibrium property (rather than a 
non-equilibrium one).
The Jarzynski identity (see the Supplementary Note 1) reads
\begin{equation}
    \label{eq:Jar}
    {\langle e^{-\beta w}\rangle=e^{-\beta \Delta F}},
\end{equation}
where $\Delta F$ is the free energy difference between the thermal
states associated with the Hamiltonians $H$ and $\tilde H$. In the absence of degeneracies, this implies
that the vector formed by the $D^2-1$ measured probabilities belongs to a
$D^2-2$ dimensional hyperplane: the ``Jarzynski manifold'' [as shown in the Supplementary Note 1, further constraints
restrict this dimensionality to $(D-1)^2$].
With the QWM we measure $P(w)$
for different driving fields showing that all probability vectors
belong to the same manifold. By
characterising this manifold, we not only verify the
identity but also independently  estimate the
free energy difference $\Delta F$ \cite{jarzynski1997nonequilibrium,hummer2001free,liphardt2002equilibrium,collin2005verification}.

The work distribution sampled by the QWM
\cite{tasaki2000jarzynski,kurchan2000quantum,mukamel2003quantum,talkner2007fluctuation} is:
\begin{equation}
    \label{eq:p-w}
    P(w) = \sum_{n,m} p_n p_{m|n} \,\delta[w - (\tilde{E}_m - E_n)].
\end{equation}
Thus, $P(w)$ is the probability density of finding the
energy difference $w$ after a
measurement of $H$ followed by an intermediate driving
$\calu_\cals$ and a final measurement of $\tilde H$. This is
indeed the case if \(p_n\) is the probability of obtaining
\(E_n\) when measuring $H$ and \(p_{m|n}\) is
the probability of obtaining \(\tilde{E}_m\) when
measuring $\tilde H$  given that \(E_n\) was detected at the beginning.
Equation ~\eqref{eq:p-w} defines a probability density that is independent of the  initial
coherences in the energy basis. For the
discrete $D^2$ values of $w$
we will use $P(w)$ to denote the
probability (not the density) of each $w$.
The concept on which our QWM is based was first discussed in Refs. \cite{roncaglia2014work,de2015measuring}, where it was noticed
that the work done on $\cals$, can be detected by  performing a generalised quantum measurement, which enables the
number of outcomes to be larger than $D$. This can be done by entangling
$\cals$ with an ancilla $\cala$ that stores a coherent record of $w$.
Then a standard measurement on $\cala$ can reveal $w$. Similar strategies have been later studied and extended to other
contexts in Refs. \cite{de2015measuring,talkner2016aspects,anders2017work}.\\\

\begin{figure*}
\includegraphics[width=1\linewidth]{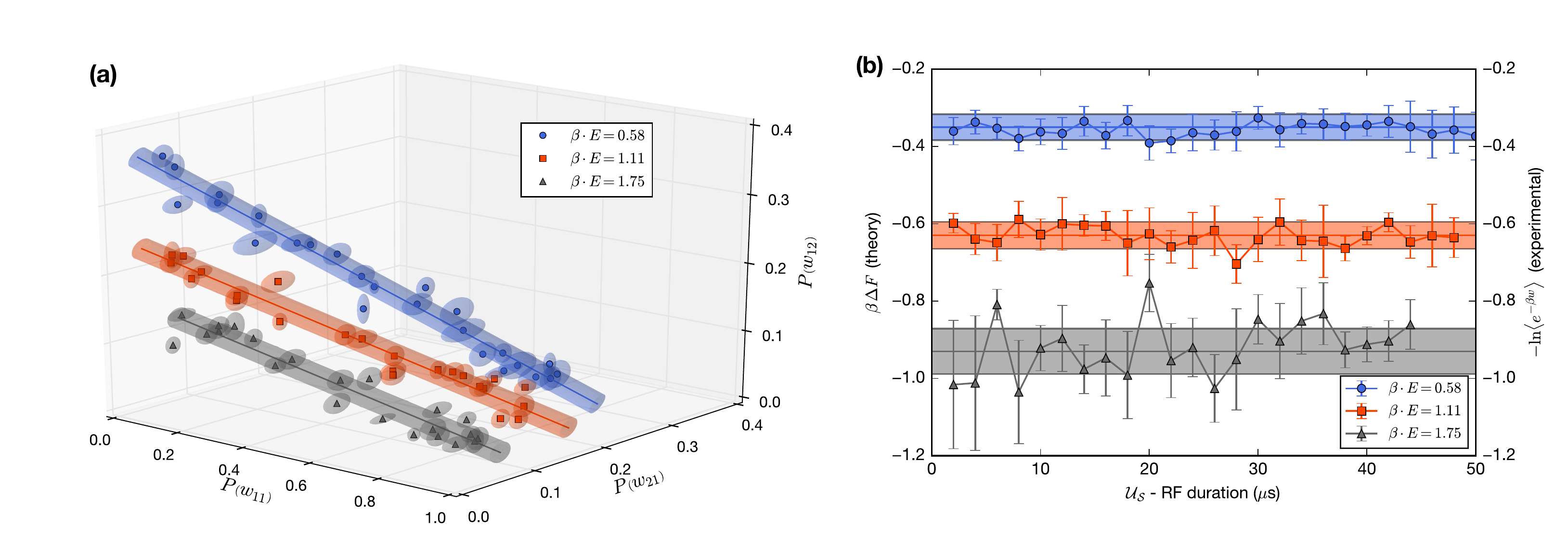}
\caption{\label{fig:3d-jarz}
The Jarzynski identity.
        {\bf(a)}~Each point defines a probability vector (with its experimental error) measured for a certain driving.
        The three lines correspond to three temperatures:
        $\beta\, E\, = 0.58\pm0.02$ (blue circle), $1.11\pm0.02$ (red square), $1.75\pm0.04$ (grey triangle).
        For each temperature all points lie in the same Jarzynski manifold (which in this case is a line).
        Reported errors are the SEM of three independent experiments with the same initial parameters and driving.
        The projections onto the three different axes of the probabilities are shown in detail in the Supplementary Figure 2.
        {\bf(b)}~${-\ln\langle e^{-\beta w} \rangle=-\ln[\sum_w e^{-\beta w}P(w)]}$ becomes independent of  the duration of the
        intermediate driving (for three  temperatures). The dots are the calculated values using
        the measured work distribution in the Jarzynski identity, and the solid line is the theoretical estimate of $\beta\Delta F$ (with an
        uncertainty due to the uncertainties in the temperature and energy splitting).
        Error bars are the SEM.
}
\end{figure*}

\noindent\textbf{Design and operation of the  QWM.}
A pictorial representation of the protocol we follow to operate the QWM is shown in Fig.~\ref{fig:esq-povm-meas}a.
The QWM is designed to measure the
work done on a system $\cals$ whose Hamiltonian
changes from $H$ to $\tilde{H}$ and which is
subjected to a driving $\calu_\cals$ in between.
We couple  $\cals$ to a
continuous variable system $\cala$
and use $\hat z_\cala$ 
to denote its position
(the generator of
translations along the momentum $p$).  A coherent record  of $w$ is created by an ``entangling interaction" between $\cala$ and $\cals$ that must take
place before and after the driving $\calu_\cals$.
The unitary operators representing these interactions are:
$U = e^{-i \lambda \,\hat z_\cala \otimes  {H} /\hbar}$
and
$\tilde U = e^{i \lambda \,\hat z_\cala\otimes {\tilde{H}}/\hbar }$,  where $\lambda$ is a coupling parameter.
Thus, $U$ and $\tilde U$ respectively
translate $\cala$ along $p$ by a
displacement proportional to $(-\lambda H)$ and $\lambda \tilde H$.
Then, as shown in detail in the Supplementary Note 3, the final measurement
of $p$ on $\cala$ yields a random result whose distribution $P_\cala(p)$ is a smeared version
of the true work distribution $P(w)$ defined in Eq.~\eqref{eq:p-w}. In fact, outcome $p$ is obtained
with a probability  density
$P_\cala(p)= \int dw P(w) f(p-\lambda w)$, where the window
function $f(p)=|\langle p|\phi\rangle|^2$ is fixed by $\ket{\phi}$, the initial state of $\cala$ [thus,
by localising  $\ket{\phi}$ we improve
the accuracy in the estimation of $P(w)$].

A ``universal'' QWM is an
apparatus which can measure $w$
and sample $P(w)$  for any possible
choice of $H$ and $\tilde H$.   To build it, we
need enough control to enforce the
entangling operators $U$ and $\tilde U$ for
any choice of $H$ and $\tilde H$.
Remarkably, this is achieved for a $2$-level system
by the atom chip implementation we
describe below.\\

\begin{table*}
    \begin{tabular}{c || c | c || c | c}
        \(\beta\, E \) & \(\beta\Delta F\) (JI) & \(\beta\Delta F\) (PF) &
        \(\Delta F/E \) (JI) & \(\Delta F /E\) (PF)  \\ \hline
        \(0.58 \pm 0.02\) & \(-0.36 \pm 0.04\) & \(-0.35 \pm 0.03\) & \(-0.62 \pm 0.07\) & \(-0.60 \pm 0.06\) \\
        \(1.11 \pm 0.02\) & \(-0.63 \pm 0.05\) & \(-0.63 \pm 0.04\) &  \(-0.57 \pm 0.05\) & \(-0.57 \pm 0.04\) \\
        \(1.75 \pm 0.04\) & \(-0.92 \pm 0.09\) & \(-0.93 \pm 0.06\) &  \(-0.53 \pm 0.05\) & \(-0.53 \pm 0.04\) \\
    \end{tabular}
    \caption{Estimates of \(\beta\Delta F\) and
        \(\Delta F\) for three different temperatures. We show the
estimation obtained using the Jarzynski identity (JI) and from a
        direct calculation of the partition function (PF).} 
            \label{tab:results}
\end{table*}

\noindent\textbf{Experimental implementation of the QWM.} 
To describe our QWM we should identify the physical systems
representing $\cals$ and $\cala$,
the way in which $H$
and $\tilde H$
can be chosen, and how the associated $U$ and $\tilde U$
are implemented. In our experiment we represent
$\cals$
by the subspace associated with the Zeeman sublevels
${\ket{1} \equiv \ket{F=2, m_F=1}}$ and $\ket{2} \equiv\ket{F=2,m_F=2}$ of a
$^{87}$Rb atom that, as in
Ref. \cite{machluf2013coherent},
behaves as a two-level system (see
below). The motional degree of
freedom of the atom plays the role of
$\cala$.

A key element of the QWM presented here is
the atom chip \cite{keil2016fifteen},
which efficiently entangles the internal and motional
degrees of freedom of an atom just
$\sim 100\,\mu$m away from the chip surface, through short and strong
Stern-Gerlach type magnetic gradient pulses.
These pulses are generated using a
$3$-current-carrying-wire setup on the chip surface
(described in Ref.~\cite{margalit2015self} and
Methods).
A gradient pulse along the $z$ direction with
amplitude $B'$ and
duration $\tau$, induces a momentum kick
$m_F\, \delta p$ on an atom in the $m_F$ state
($\delta p\sim\mu_B g_F B'\tau$, where $\mu_B$
and $g_F$
are, respectively, the Bohr magneton and the Land\'e factor\,\cite{machluf2013coherent}).
The evolution of the state of the atom induced by such
a pulse is described by the unitary operator
$U_p=e^{i \,\delta p\,\hat z_\cala \otimes \hat \sigma/\hbar}$,
where the operator  ${\hat\sigma=\ket{1}\bra{1}+2\ket{2}\bra{2}}$.
This physical operation translates $\cala$ along the
momentum $p$ by a displacement $\delta p
\, \hat \sigma$ (notice that the operator  
$\hat\sigma$ defines the magnetic dipole moment
of the atom since $\hat\sigma=\sum_{m=1,2}m
\ket{m}\bra{m}$).
As described below, we apply two gradient pulses with
different amplitudes ($B'$ and $\tilde B'$) and
different durations ($\tau$ and $\tilde\tau$).
Thus, defining $H=E\hat{\sigma}$  and ${\tilde H}={\tilde E}\hat{\sigma}$,  $U_p$  and $\tilde{U}_p$ implement
the required entangling operation $U$ and $\tilde{U}$, respectively.
In this implementation $\lambda$ is consequently replaced
by $-\delta p/E$ and $\delta\tilde p/\tilde E$,
enforcing $\tilde E / E = -\delta \tilde p/\delta p$.
The momentum kicks induced by both
pulses are controlled in the experiment, and consequently, by fixing their ratio, we can simulate an arbitrary system
with initial and final Hamiltonians $H$ and $\tilde H$ which are characterised by $\tilde E / E$ having the same ratio.
Finally, let us note that the two pulses utilise $B'$ and $\tilde B'$ with opposite signs to ensure that
the sequence creates a record of work corresponding to $\tilde E_m-E_n$.

To achieve universality we only need to
be able to fix the energy splitting $E$ and $\tilde E$ of $H$ and
$\tilde H$, as well as their eigenbasis. 
The traces of $H$ and $\tilde H$ (the sum of
their eigenvalues) do not affect
$P(w)$ but only add a constant to all values
of $w$. As arbitrary $E$ and $\tilde E$ can be
simulated and any change of basis can be absorbed into $\calu_\cals$, we
conclude that our atom chip QWM can sample $P(w)$ for an arbitrary 2-level system and is thus universal.

The $^{87}$Rb atoms are magnetically
trapped in state \(\ket{2}\) and evaporatively cooled to a Bose-Einstein condensation (BEC).
The BEC is released from the trap and a radio-frequency (RF) pulse is used to prepare a  superposition of $\ket{1}$
and $\ket{2}$. A strong
homogeneous magnetic
field (created by external coils)
suppresses the transitions 
taking $\ket{1}$  into the $\ket{2,0}$ state
(due to the
non-linear Zeeman effect  \cite{machluf2013coherent}).
The initial populations ($p_1$
and $p_2$) are chosen so that  $\beta E = \ln{(p_1 / p_2)}$.
The initial motional state is a wave-packet localised in position and momentum.

It should be noted that the initial internal 
state of the atom, while having the same
populations as defined by the temperature of a thermal state, is 
still a pure state. However, 
the quantum coherences of this initial state
do not affect 
the results of the QWM. As explained in 
the Supplementary Note 3, the contribution of the initial 
coherences to the final probability is 
multiplied by the overlap between the 
motional states of the atom associated 
with the different values of work. Thus, 
when the atomic clouds associated with 
the different work values are 
well separated, the effect of initial coherences
is negligible. In this regime, our experiment 
gives the same result as the one we would 
obtain by preparing an initial thermal  state
(with no coherences). The study of the
importance of the initial coherences in the
definition of work is an interesting topic in 
itself, which is beyond the scope of our paper 
(see, for example,  
Ref. \cite{solinas2015full,kammerlander2016coherence,perarnau2017no,anders2017work,sampaio2017impo}).

The experimental sequence, presented in
Fig.\,\ref{fig:esq-povm-meas}b,
is: i)~prepare the initial state
and release the cloud (which then freely falls along $z$, the
direction of gravity), ii)~apply the magnetic gradient $U$ along $z$, iii)~apply
the driving $\calu_\cals$ by exposing
the atoms to a RF field resonant with the
Zeeman splitting induced by the homogeneous bias field,
iv)~apply the gradient $\tilde U$,
v)~obtain an image of the four clouds after a time-of-flight
and count the number of atoms in each cloud.
More details of the experiment can be found in Methods and the Supplementary Note 3.
For the experimental demonstration presented here 
we set the ratio between the
measured momentum kicks induced by the two pulses to $-\delta \tilde p/\delta p=0.56\pm0.02$. 
Hence, our realisation of the QWM samples the work distribution of a simulated system in which the energy splitting is reduced to $56\%$
of its original value, from $E$ to $\tilde E$, while driven by $\calu_\cals$. 

Fig.~\ref{fig:esq-povm-meas}c shows a typical image
obtained by the QWM. Four clouds are
visible. 
From the positions of 
the center of each cloud, $\bar z$, we infer 
the total momentum shift, $\bar p$,  
induced by the pulses on that cloud 
(we take into account both the  
free fall and the kicks induced by the pulses, see the Supplementary Note 3). 
Then, we obtain the corresponding value of work 
as $w=E \, \bar p/\delta p$ ($w$ is 
proportional to $E$, whose value, together with 
the experimental results, determine the work $w$).  
Furthermore, the probability $P(w)$ for each $w$ is directly measured 
by the number of atoms in each cloud. 
Notably, this experiment determines the entire $P(w)$ distribution
in a single shot.\\

\noindent\textbf{Testing the Jarzynski identity.}
We repeat the experiment fixing the
timing, duration and pulse
strength. We consider
three initial $\beta$'s and vary the intermediate driving $\calu_\cals$ by
changing the duration of the RF field.
In this way, we obtain many sets of  probability
distributions, each of which defines
a 3D--vector (as there are three independent probabilities). When we
represent all
these vectors in the same 3D-plot, we
see that they all belong to the
same $\beta$--dependent manifold.
Fig.~\ref{fig:3d-jarz}a shows that
this manifold is a $\beta$--dependent line
(the dimensionality of this ``Jarzynski manifold"
is $(D-1)^2$, which in this case equals $1$).

Using the measured work probabilities
we calculate the exponential average
of the work \(\langle e^{-\beta w} \rangle\) for each
driving field.
Fig.~\ref{fig:3d-jarz}b displays the value of
${G=-\ln[\langle e^{-\beta w}\rangle]=-\ln[\sum_w  e^{-\beta w} P(w)]}$
as a function of the duration of the intermediate RF field,
that parametrises $\calu_\cals$. As established by the
Jarzynski identity, $G$ is independent of the driving
field and only depends on
$\beta$. The horizontal lines in Fig.~\ref{fig:3d-jarz}b
are the theoretically predicted values
of $\beta\, \Delta F$, obtained 
from a direct calculation (with its own theoretical uncertainty, due to the error
 in the estimation of $\beta\, E$). This calculation simply
involves computing the initial and 
final partition functions, respectively 
denoted as 
$Z$ and $\tilde Z$, and using the identity 
$\beta\,\Delta F = \ln(Z/\tilde Z)$.
We find that, as
 the Jarzynski identity establishes, $G=\beta\Delta F$.
From Fig.~\ref{fig:3d-jarz}b one can notice that the largest errors in the estimation of $\beta\Delta F$ appear for $\beta \, E =  1.75$.
In this case, $P(w)\lesssim 0.1$ for two values of $w$ and, therefore,
the relative error in the atom number estimation is large, inducing a larger error in the estimation of $\beta\Delta F$.

In Table \ref{tab:results} we compare
measured and estimated values of
$\beta\Delta F$. The uncertainty in the estimation of
\(\beta\Delta F\) and \(\Delta F\) is close to
10\%, which is enough to distinguish the three values of \(\beta\Delta F\).
On the other hand, in the case of $\Delta F$, there is a significant
overlap in the measured values which does not allow to properly distinguish between the three
different cases due to the error in the estimation of $\beta E$.\\

\section{Discussion}
We presented and implemented 
a QWM, a new device directly sampling
the work distribution on an ensemble of
cold atoms. Our QWM can be used to simulate
the behaviour of an arbitrary $2$-level system. 
We implemented it with an atom chip and verified the
Jarzynski identity over a wide range of non-equilibrium processes. This is the first experiment,
and so far the only one, directly sampling $P(w)$
offering advantages and
different perspectives over previous work measurement schemes.
Remarkably, in this cold  atom experiment, the QWM
 extracts full statistical information
about the work distribution in a single shot. \\

\section{Methods}

\noindent\textbf{Initial state preparation.}~ After preparing the BEC,
a homogeneous magnetic field of $36.7\,$G ($25\, h\,\mathrm{MHz}/\mu_B$, where $h$ is Planck's constant) is used to 
push the transition to $\ket{2,0}$ out of resonance by
$\sim 180\,$kHz due to the
non-linear Zeeman effect, which is larger than the power
broadened driving RF field of $\calu_\cals$. This ensures that the atoms 
behave as $2$-level systems.
The BEC is released from the trap and a RF pulse is used
to prepare a superposition of $\ket{1}$
and $\ket{2}$. By varying the relative populations we consider three different pseudo-thermal
states.
The initial motional state is a wave packet $\ket{\phi}$,
well localised at \(z_0= 91\pm 1.2\,\mu\)m  
from the chip
and momentum $\sim0$.\\

\noindent\textbf{Entangling operations and measurement.}~An
inhomogeneous magnetic field is used to couple spin and motional
degrees of freedom.
This is generated by a current $I= 0.85\,$A in the 3-wire setup
during a time $\tau$. 
The three parallel gold wires lie on the $x$ direction
of the chip surface (Fig. 1.b). They
are $10$ mm long, $40\, \mu$m wide and $2\, \mu$m
thick. Their centers are at $y=-100, 0, 100\, \mu$m 
and the same current run through them in 
alternating directions ($-I,I,-I$, respectively),
creating a $2$D quadrupole field at $z=100\, \mu$m
below the chip.  
After a time of flight of $2.4$\,ms the atoms are at
$z\sim 119\,\mu\mathrm{m}$. At this point the
first gradient pulse implements $U$:
$\tau = 40\,\mu$s with an amplitude of $B'\sim 95\,$G/mm,
such that the momentum kick is along $+z$.
Then, after 3.1~ms the atoms are at $\tilde z \sim 0.3\,$mm
and the second gradient pulse implements $\tilde U$:
$\tilde \tau = 300\,\mu$s,
$\tilde B'\sim -7.5\,$G/mm, such that the momentum
kick is along $-z$. The relative strengths of the
spin-dependent forces sets the energy splitting of
the Hamiltonians which in this case is on average
$\tilde E/E=-\tilde \delta p/\delta p=0.56\pm0.02$ 
(this is the measured value, that takes into account
fluctuations in the initial
position of the cloud and in the gradient pulses).
In between the entangling operation $\calu_\cals$ is applied
with a RF pulse.
Finally, an image of the atomic clouds is obtained after a 
time-of-flight of $18.2\,$ms after the $2$nd gradient 
(the clouds are centered around $z\sim 3 \,$mm). The
position and
number of atoms of each cloud is determined.
The momentum shifts of each cloud (that codifies the value of $w$)
are obtained from the difference in positions between the
clouds, that follow approximately classical trajectories (see Supplementary Note 3). \\

\noindent\textbf{Uncertainties.}
The main source of position error is the initial distance of the cloud from the atom chip, whose uncertainty is $\sim 1\%$.
This error is later translated to momentum uncertainty, since the field gradients are position dependent.
The  field gradients have a
fractional uncertainty of $10^{-3}$ due to current fluctuations \cite{machluf2013coherent}.
The central position of each
cloud is estimated by fitting a Gaussian profile.
Each work probability is estimated as a normalised sum of the measured
optical density in a relevant region around the cloud,
introducing probability uncertainty (due to atom numbers uncertainty). Our $\sim 5\,\mu\mathrm{m}$ optical resolution also induces an error in the determination of the position for each cloud. 
We perform three different runs for each combination of initial 
state population ratios and intermediate driving
and use the average values of position and probability.
This gives us a position uncertainty of $\sim 0.015\,\mathrm{mm}$ and a probability uncertainty
of $\sim 0.015$ (standard error).\\

\vspace{1cm}
\noindent
\textbf{Acknowledgments.}
FC, AJR and JPP acknowledge financial support from ANPCyT (PICT 2013-0621 and PICT 2014-3711), CONICET
and UBACyT. YM and RF gratefully acknowledge funding by the
Israel Science Foundation, the EC Matter-Wave consortium
[FP7-ICT-601180],
and the German DFG through the DIP program [FO703/2-1]. SM acknowledges
financial support by the Foundation for Fundamental Research on Matter (FOM),
which is part of the Netherlands Organisation for Scientific Research (NWO). 
We also thank the BGU nano-fabrication team for
making available the high quality chip, and Zina Binstock
for helping with the electronics. \\

\newpage

\onecolumngrid
\appendix

\section{Supplementary Material.}

\section*{Supplementary Note 1 - Jarzynski identity and quantum work measurement}
\label{ap:Jarzynski}
Work is the energy variation induced on a system 
$\cals$ by a certain driving (the 
system is otherwise isolated). Suppose that the 
Hamiltonian of $\cals$ changes from an initially 
$H$ to a final $\tilde H$ and that the system is driven
by a certain unitary operator $\calu_\cals$ in between. 
We denote the eigenvalues and eigenvectors of 
$H$ as $E_n$ and 
$\ket{\varphi_{n,\alpha}}$ (where $\alpha$ labels 
different eigenstates with the same
energy). In turn the eigenvalues and eigenvectors of $\tilde H$
are denoted as $\tilde E_m$ and 
$\ket{\tilde\varphi_{m,\gamma}}$ (again, $\gamma$ labels
states with the same energy $\tilde E_m$). 
Suppose that we measure the energy at the initial and final 
times. The work probability distribution $P(w)$ is
nothing but the probability density to obtain a value $w$ as 
the difference between the results of the two energy 
measurements. This can obviously
be computed as
\begin{equation}
    \label{eq:p-w}
    P(w) = \sum_{n,m} p_n p_{m|n} \,\delta\left[w - (\tilde{E}_m - E_n)\right],
\end{equation}
where \(p_n\) is the probability of initially 
measuring energy \(E_n\) and \(p_{m|n}\) is 
the probability of measuring energy
\(\tilde{E}_m\) at the end of the driving 
given that \(E_n\) was detected at the beginning.
If the initial state of $\cals$ is $\rho$, the above probabilities
can be simply written in terms of the projectors 
$\Pi_n=\sum_\alpha \ket{\varphi_{n,\alpha}}\bra{\varphi_{n,\alpha}}$ and
$\tilde \Pi_n=\sum_\gamma \ket{\tilde\varphi_{n,\gamma}}\bra{\tilde \varphi_{n,\gamma}}$. Thus,  
\beq
p_n=\tr(\rho\,\Pi_n), \qquad p_{m|n}=\frac{1}{p_n}
\tr(\tilde\Pi_m \calu_\cals\Pi_n\rho\,\Pi_n\calu^\dagger_\cals).
\label{eq:prob}
\eeq
Jarzynski identity follows immediately from the 
above definition of the work probability distribution. 
In fact, if we compute the exponential average of the work, 
we get
\begin{eqnarray}
\langle e^{-\beta w}\rangle&=&\int dw P(w) e^{-\beta w}\nonumber\\
&=& \sum_{n,m} p_n p_{m|n} e^{-\beta(\tilde E_m-E_n)}.
\end{eqnarray}
If the initial state is thermal, then $\rho=
\sum_n\Pi_n e^{-\beta E_n}/Z$ (where $Z$ is
the partition function $Z=\sum_n g_n e^{-\beta
E_n}$ with  $g_n=\tr(\Pi_n)$ being the degeneracy of level $E_n$). Then, 
we can replace  $p_n=e^{-\beta E_n}/Z$ and perform the summation over the label $n$ by noticing that
$\sum_n p_{m|n}=\tr(\tilde \Pi_m)=\tilde g_m$. 
In this way we obtain
\begin{equation}
\langle e^{-\beta w}\rangle={1\over{Z}}
\sum_{m}\tr(\tilde\Pi_m) 
e^{-\beta \tilde E_m}=\frac{\tilde Z}{Z},\nonumber
\end{equation}
where the partition function of the final Hamiltonian is $\tilde Z=\sum_m \tilde g_m e^{-\beta\tilde E_m}$, with
$\tilde g_m=\tr(\tilde\Pi_m)$ the degeneracy of level $\tilde E_m$. 
The final step  that leads to the Jarzynski identity is simply to notice that for a Gibbs state we 
have $e^{-\beta \Delta F}=\tilde Z/Z$. Thus, 
the average exponential work becomes independent of the intermediate driving process $\calu_\cals$ and turns out
to be determined by equilibrium properties. Thus, 
$\langle e^{-\beta w}\rangle=e^{-\beta\Delta F}$. 

When $\cals$ has a finite dimensional Hilbert space, 
$w$ can only take discrete values. In this case, instead of
using the probability density we use directly the 
probability for each value of $w$. Thus, we can 
arrange $P(w)$ as a vector with $D^2$ components. Taking into account the 
normalisation condition, the vector of 
independent probabilities is $D^2-1$ 
dimensional. In turn, being the Jarzynski 
identity a linear equation in terms of $P(w)$
it constraints the probability vector to belong
to a $D^2-2$ dimensional hyperplane. 
But, of course, there are further constrains 
reducing the number of independent probabilities. 
The simplest way to obtain 
this number is to go back to 
the definition of $P(w)$ in Supplementary Equation~\eqref{eq:p-w} 
and count the number of free parameters we 
have. In this 
equation, the probabilities $p_n$ are fixed by the initial 
state. Then, the number of independent probabilities is
determined by the number of independent parameters in $p_{m|n}$.  For the non-degenerate
case we consider here (where all the values of work are different), the calculation is simple. 
In fact, the coefficients $p_{m|n}$ 
form a doubly stochastic matrix (since they 
are all positive numbers such that $\sum_np_{m|n}=\sum_mp_{m|n}=1$). For such square matrix
of dimension $D$, there is always 
$(D-1)^2$ free parameters. Indeed, this is the dimensionality of the manifold of where the 
probability vector lies. Jarzynski identity
establishes that this manifold is a $\beta$-dependent hyperplane (a line in our case, where
$D=2$). 

\section*{Supplementary Note 2 - Work measurement as a POVM}
\label{ap:povm}
Let us consider a system $\cals$ with a $D$-dimensional
space of states and show, in a simple way, that the 
work measurement can be viewed as a generalised quantum
measurement. For this, we start by 
writing the probability for a given value of work 
$w_{nm}=\tilde E_m-E_n$ as 
\beq
P(w_{nm})= p_{m|n}\, p_n.
\eeq
Using the formula for the transition probability, we
find that
\beq
P(w_{nm})= \tr(\tilde\Pi_m \calu_\cals\Pi_n\rho\Pi_n\
\calu_\cals^\dagger)=\tr(\rho A_{nm}), 
\eeq
where 
\beq
A_{nm}=\Pi_n\calu_\cals^\dagger
\tilde\Pi_m \calu_\cals\Pi_n. 
\eeq
It is simple to verify that the operators  $A_{nm}$ 
expand the identity as $\id=\sum_{n,m} A_{nm}$ and that they
are positive semi-definite (i.e., that for any state $\ket{\chi}$
we have $\bra{\chi}A_{nm}\ket{\chi}\ge 0$). Therefore, 
the operators $A_{nm}$ define a positive operator
valued measure (POVM), which is the most 
general type of measurement one can perform
in quantum mechanics. 
Therefore, work can be measured in the same
way as any POVM can: A powerful 
result (Neumark's theorem) establishes that 
any POVM can be realised 
by coupling the system $\cals$ with an ancillary
system $\cala$ and then performing a standard 
projective measurement
on $\cala$. This measurement can be 
performed at a single time. Thus, 
surprisingly, the two-time work measurement
strategy can be replaced by a single-time 
strategy (which is the basic idea exploited by our QWM). 
In the following section we show how one can construct an approximation for that ideal apparatus, that we call Quantum Work Meter (QWM).

\section*{Supplementary Note 3 - Probability distribution for the outcome of a QWM}
\label{ap:prob}
Here we compute the probability distribution for the 
result of the measurement of the auxiliary register of a 
general Quantum Work Meter. 
The protocol defining the apparatus is shown in Figure 1 of the main text. A system $\cals$
is coupled to an ancillary one $\cala$. This ancilla 
is a continuous variable system (of course this can be
relaxed). The system $\cals$ and the ancilla $\cala$ are subject to 
the following evolution: $i)$ an entangling interaction $U$
is applied (which correlates $\cals$ and $\cala$), 
$ii)$ the evolution $\calu_\cals$ is applied on the system
$\cals$, 
$iii)$ a second entangling interaction $\tilde U$ is
applied. Finally, after this sequence $\cala$ is measured.
The initial state of the system formed by 
$\cals$ will be assumed to be a product 
state. For simplicity, we first assume that 
the states are pure and denote them 
as $\ket{\xi}$ (the state of $\cals$)
and $\ket{\phi}$ (the state of $\cala$). We 
will later generalise the result for an 
initial state which is a tensor product of
arbitrarily mixed states. 
After the sequence of operations we described
above, the total final state is: 
\begin{equation}
\ket{\Phi(t_f)}=
\tilde U\, (I_\cala\otimes\calu_\cals )\,
U\; \ket{\phi}\otimes\ket{\xi}\nonumber
\end{equation}
The nature of the two entangling 
operations, that was discussed
in the main text, is such that they both 
induce translations of $\cala$ which depend
on the state of $\cals$. More specifically,   
 $U=e^{-\frac{i}{\hbar}\lambda \hat z_\cala \otimes H}$ and 
 $\tilde U=e^{\frac{i}{\hbar}\lambda \hat z_\cala \otimes \tilde H}$, 
where we use $\hat z_\cala$ to denote 
the generator of 
translations of $\cala$ along a certain variable $p$. 
Using this, it is simple to rewrite the final state as
 \begin{equation}
 \ket{\Phi (t_f)}=
 \sum_{n,m} D_{nm}\ket{\phi} \otimes  \tilde\Pi_m\calu_\cals\Pi_n\ket{\xi} 
             \nonumber
 \end{equation}
 In this equation the displacement operators $D_{nm}$ 
 act on the states of $\cala$ and 
 are defined as $D_{nm}=e^{ \frac{i}{\hbar} \lambda w_{nm}\hat z_\cala}$. 

The interpretation of the above equation is simple: After the sequence of operations, 
$\cals$ and $\cala$ become entangled 
in such a way that a record of $w_{nm}$ is stored
in  $\cala$. The states $\ket{F_{nm}}\equiv
D_{nm}\ket{\phi}$ are ``flag states" associated
with the different values of work. When these
states are orthogonal, they can be 
unambiguously distinguished and 
the value of work can be 
retrieved.  Below, we will consider a more 
realistic scenario where the initial state of 
$\cala$ is a localised coherent
state. In that case, the flag states are 
displaced coherent states (which are simply
translated along the variable $p$ direction by 
an amount that is proportional to $w_{nm}$). 
These states are not 
strictly orthogonal, but have a 
finite overlap. This induces
an error in the work estimation protocol. 
However, the error can be exponentially
reduced by simply 
increasing the interaction strength 
$\lambda$ (as the overlap exponentially 
decreases with $\lambda$). 

From the above expression it is simple to obtain 
the quantum state of $\cala$ by computing
its reduced density matrix (which 
is obtained from the total state by tracing out the system $\cals$). Thus, 
\beq
\rho_\cala(t_f)=
\sum_{n,n',m} \tr\left(\tilde\Pi_m
\calu_\cals\Pi_n\ket{\xi}\bra{\xi}
\Pi_{n'}\calu^\dagger_\cals\right)\;
D_{nm} \ket{\phi}\bra{\phi} D^\dagger_{n'm}.
\eeq
This expression can be generalised to the
case where the initial states of $\cals$ and 
$\cala$ are initially mixed. In fact, if  
$\rho_\cals$ and $\rho_\cala$ respectively
denote the initial density matrices of $\cals$ and 
$\cala$, the final state of $\cala$ is
\beq
\rho_\cala(t_f)=
\sum_{n,n',m} \tr\left(\tilde\Pi_m
\calu_\cals\Pi_n\rho_\cals\Pi_{n'}\calu^\dagger_\cals\right)
D_{nm} \rho_\cala D^\dagger_{n'm}.
\eeq
From the above equation we obtain the 
probability density for detecting the value $p$
in a measurement of $\cala$. 
Thus, 
\beq
P_\cala(p)=
\sum_{n,n',m} \tr\left(\tilde\Pi_m
\calu_\cals\Pi_n\rho_\cals
\Pi_{n'}\calu^\dagger_\cals\right)
\bra{p-\lambda w_{nm}}\rho_\cala\ket{p-\lambda w_{n'm}}.
\eeq
The contribution of the diagonal ($n=n'$) and 
off-diagonal ($n\neq n'$) terms play a different
role in the above expression. 
In fact, it is simple to show that 
the diagonal contribution is a 
smeared version of the true work distribution. 
Thus, 
\beq
\sum_{n,m} \tr \left(\tilde\Pi_m
\calu_\cals\Pi_n\rho_\cals
\Pi_{n}\calu^\dagger_\cals\right)
\bra{p-\lambda w_{nm}}\rho_\cala\ket{p-\lambda w_{nm}}=\int dw \,P(w) \, f(p-\lambda w), 
\eeq
where the window function is
\beq 
f(p)=\bra{p}\rho_\cala\ket{p}.
\eeq
Therefore, the off-diagonal terms of $\rho_\cals$ are responsible 
for the error in the work 
estimation and should be made 
small for it to be accurate.  It is simple to show that if the momentum 
wave function of the initial state $\ket{\phi}$
is a Gaussian with a momentum dispersion $1/\sigma$ ($\sigma$ is the position dispersion), 
then the off-diagonal terms are bounded by:
\beq
\left |\sum_{n\neq n',m} 
\tr \left (\tilde\Pi_m
\calu_\cals\Pi_n\rho_\cals
\Pi_{n'}\calu^\dagger_\cals \right ) 
\braket{p-\lambda w_{nm}}{\phi}\braket{\phi}{p-\lambda w_{n'm}} \right | 
\leq \sum_{n\neq n',m} 
\left |\tr\left (\tilde\Pi_m
\calu_\cals\Pi_n\rho_\cals
\Pi_{n'}\calu^\dagger_\cals \right ) \right | \,
\frac{\sigma}{\hbar \sqrt{\pi}} e^{-\frac{\sigma^2\lambda^2}{4 \hbar^2}(E_n-E_{n'})^2}
\nonumber
\eeq
Thus, by increasing $\lambda$ (the interaction 
strength) or $\sigma$ (the position dispersion
of the initial state) we exponentially reduce
the error in the work estimation. It is 
worth noting that by increasing $\sigma$
we reduce the momentum uncertainty and
localise the initial state in 
momentum. Naturally, the method becomes
precise when the initial localisation in momentum is much smaller than the difference between the first momentum kicks (which is fixed by 
the product $\lambda(E_n-E_{n'})$).
When these conditions are satisfied, the 
off-diagonal terms can be neglected and 
the probability density to detect $p$ is 
\begin{equation}
  P_\cala(p)=\int dw \, P(w)\, f(p-\lambda w),\nonumber
\end{equation}
where the window function $f(p)$ is defined as $f(p)=|\braket{p}{\phi}|^2$ (which, for a coherent
state is simply ${f(p)=e^{-\frac{\sigma^2}{\hbar^2}p^2} \sigma/\hbar \sqrt{\pi}}\;$). 

\subsubsection*{QWM using an atom chip}
\begin{figure}[htbp]
    \includegraphics[width=0.35\textwidth]{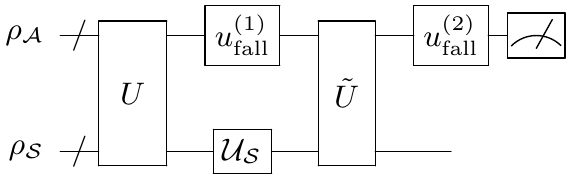}
\caption{Set of gates describing the atom 
chip QWM. $u^{(1)}_{\rm fall}$ and $u^{(2)}_{\rm fall}$ are the free fall evolution that the atoms feel during the experiment. Finally there is a measurement of the position of the atoms.}
\label{fig:circuit}
\end{figure}

Here we consider the implementation of the QWM  using a cloud of atoms in a chip. As 
described in the main text, 
$\cals$ is the pseudo spin 1/2 associated with the $F=2$, $m_F=1,2$ hyperfine states of a 
$^{87}\rm{Rb}$ atom (which, as discussed in the main text, behaves
as a two level atom). $\cala$ is encoded in the motional degrees
of freedom of each atom. There are two subtle
differences between the ideal protocol 
for a QWM described in the previous section
and the implementation in an atom chip. 
The first different concerns the final 
measurement. Thus, in the previous section 
we computed the probability for a final 
momentum measurement but in the real 
experiment we are forced to measure the 
atomic position by taking an 
image of the atomic clouds. Therefore we
will show below that the position measurement
enables us to determine work and sample $P(w)$. 
The second difference is that in the real 
experiment we should take into account the
fact that the ancilla $\cala$ evolves 
during the whole process because the 
atoms actually move (they 
freely fall along the vertical direction). 
Thus, the real protocol describing the experiment
is shown in Fig. \ref{fig:circuit}. 

Let us now analyse this process. We can first
neglect the free fall taking place between 
the two entangling operations (we take this 
into account later) and assume the initial state 
of $\cala$ is a coherent state localised around
initial values of position and momentum which 
we arbitrarily take as $z=0$ and $p=0$. 
We denote this state as $\ket{\phi}=\ket{0,0}$. 
The calculation presented in the above section
should be slightly modified. In this case the 
flag states are $\ket{F_{nm}}=u^{(1)}_{fall}
D_{nm}\ket{0,0}$. Taking into account that 
$u_{fall}=e^{-\frac{i}{\hbar} t(\hat p_\cala^2/2 m_a-m_a g \hat z_\cala)}$ (where 
$t$ is the duration of the free-fall, $m_a$ is the mass of the atoms, and $g$ the gravity acceleration) we
can easily compute the expectation value of
the position for each flag state (as well as the
corresponding position dispersion). In fact, 
we have 
\beq
z_{nm}(t) =- w_{nm}\,\frac{\lambda t}{m_a}+\frac{g}{2}t^2  \qquad {\rm and} \qquad  \Delta z_t=
\frac{\sigma}{\sqrt 2} \sqrt{1+ \left (\frac{\hbar\, t}{m_a \sigma^2}\right)^2}.
\eeq
Therefore, we can notice that by measuring the final 
position of the atoms we can infer the value of the momentum before the fall 
and thus acquire information about  work $w$. In fact, the difference between the positions of the
clouds is proportional to the difference in the values of work. The price we have to pay, is that the spread of the 
wave packets increases during the free-fall. 

In a more realistic description of 
the experiment, we need to include also the free fall between the entangling gates ($u_{\rm fall}^{(1)}$ in the Supplementary Figure  \ref{fig:circuit}). 
It is easy to verify that since 
$u_{\rm fall}^{(1)} \,\hat z_\cala \, u_{\rm  fall}^{(1)\dagger} = \hat z_\cala + \hat p_\cala\, t/m_a  + \id_\cala \, g \, t^2/2$ (where $t$ is the duration of the free fall), then
$u_{\rm fall}^{(1)} \,e^{\frac{i}{\hbar}\lambda\, \hat z_\cala\otimes \tilde H} \, u_{\rm fall}^{(1)\dagger} =
e^{\frac{i}{\hbar} \lambda\, \hat z_\cala \otimes \tilde H} 
e^{\frac{i}{\hbar}\frac{t}{m} \lambda\, \hat p_\cala \otimes \tilde H} e^{\frac{i}{\hbar}\theta\, \id_\cala\otimes \tilde H}$.
Thus, the first term is the usual entangling operation and the last term is just a phase depending on the value of the energy. In 
turn, the second term is an entangling operation
where the atom is displaced along position (instead of momentum) depending on the state of $\cals$. In summary, 
the free-fall in between the entangling gates simply induces an extra 
translation of the atoms by an amount that  depends on the final value of the energy.

Finally, we show in the Supplementary Figure \ref{fig:projections} the distribution probability that we obtain in each experiment.
In the plots we show the projections of the Jarzynski manifold, that appears in Fig. 2 (a) of the main text, onto the different axes.
\begin{figure*}
    \begin{minipage}[b]{0.4\textwidth}
   \centering
        \includegraphics[width=1\linewidth]{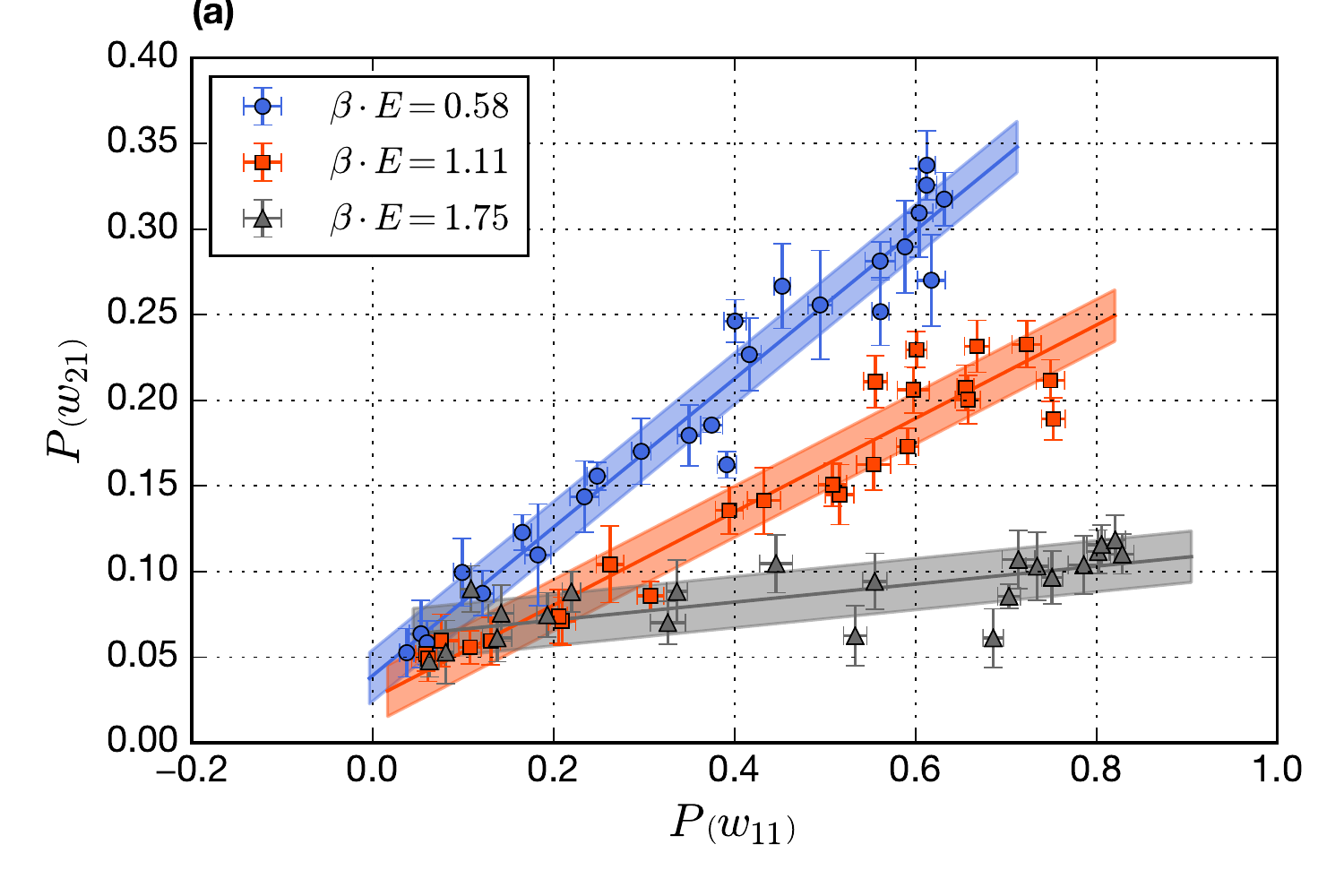}
   \end{minipage}
   \begin{minipage}[b]{0.4\textwidth}
        \includegraphics[width=1\linewidth]{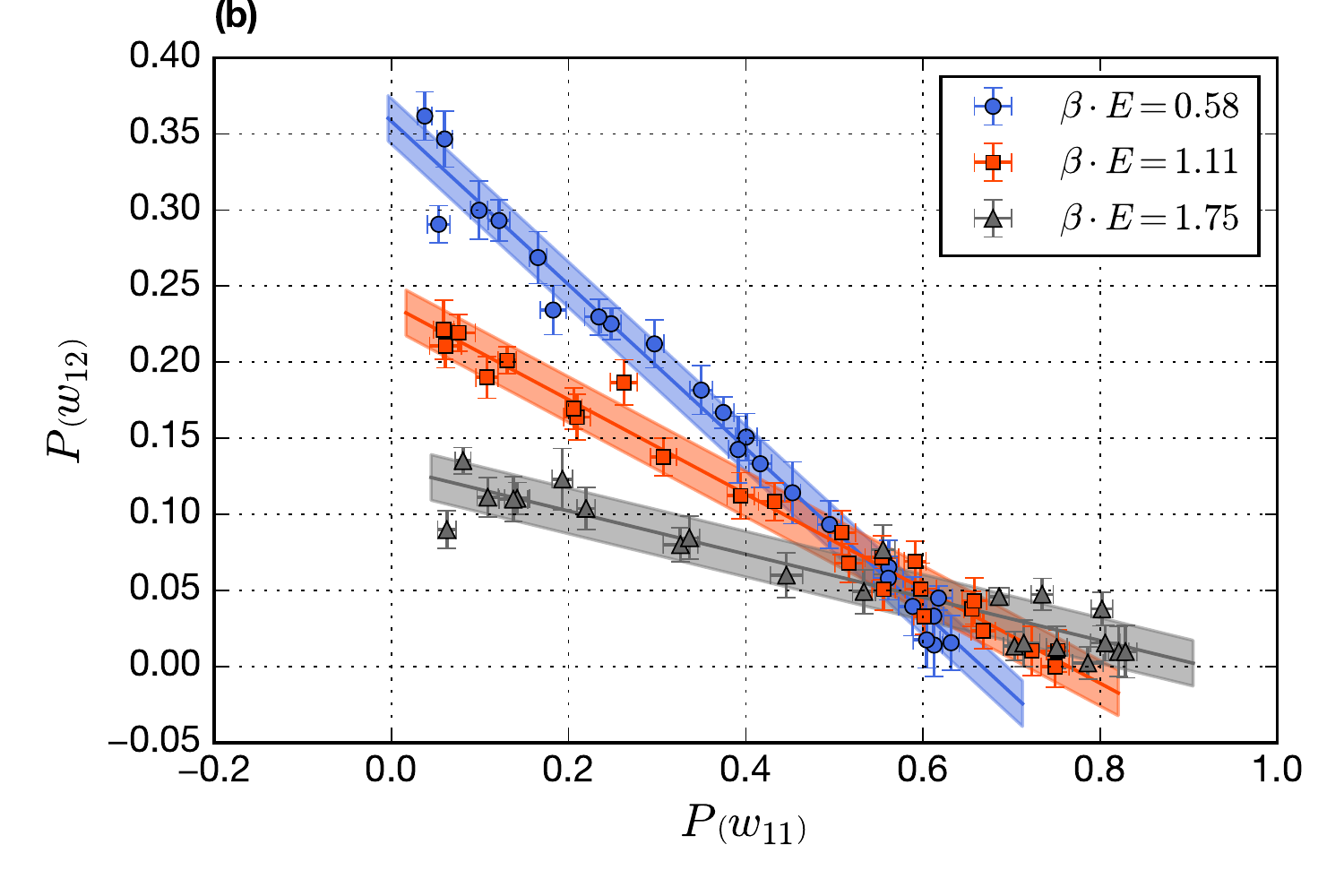}
   \end{minipage}
   \begin{minipage}[b]{0.4\textwidth}
        \includegraphics[width=1\linewidth]{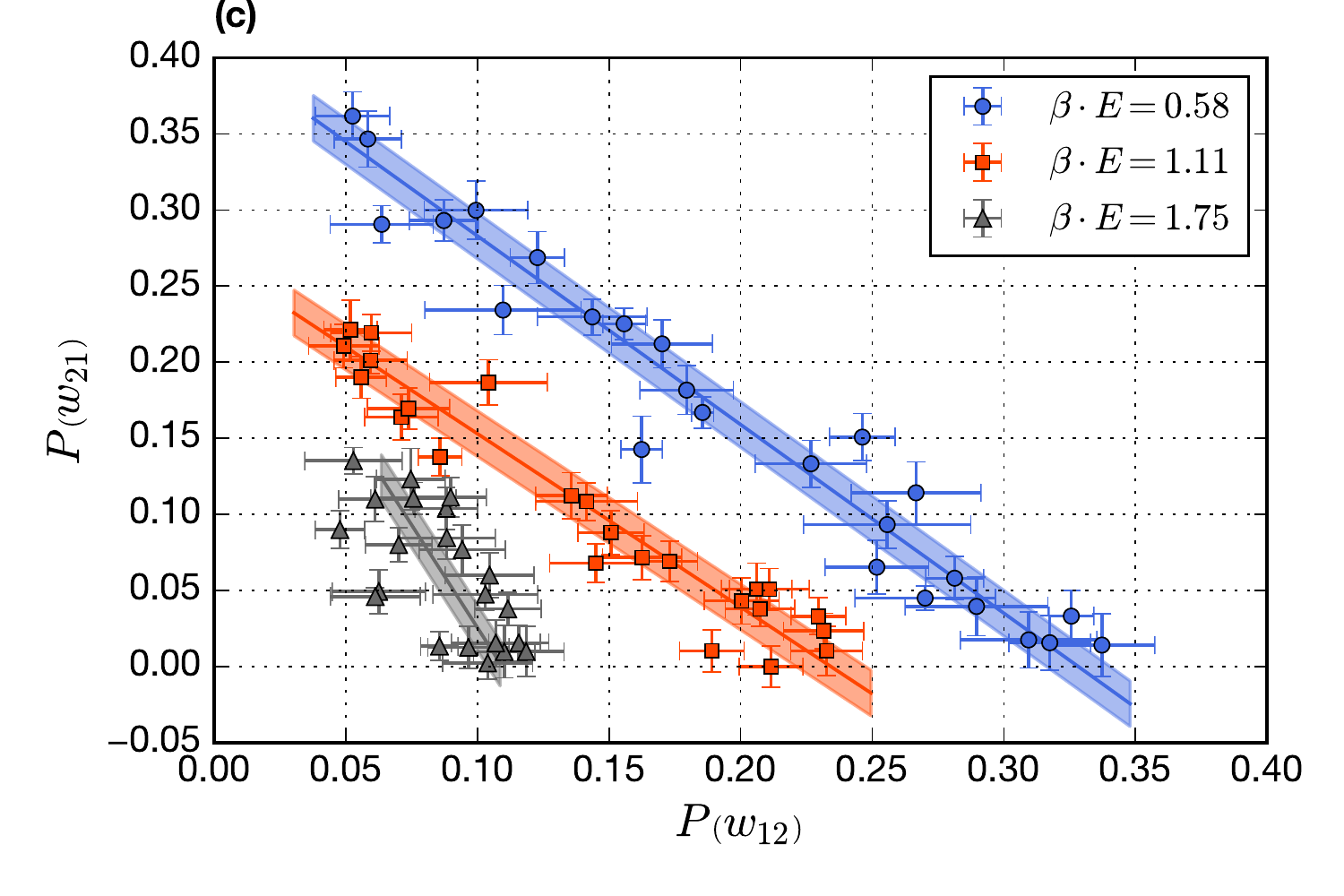}
   \end{minipage}
\caption{\label{fig:projections}
Measured work probabilities. {\bf(a)--(c)} Each point defines a probability vector (with its experimental error) measured for a certain driving. 
Error bars are the SEM of three independent experiments.
The plots are the projections of the Jarzynski manifold, that appears in Fig. 2 (a) of the main text, onto the different axes.
The three lines correspond to three temperatures: $\beta\, E\, = 0.58\pm0.02$ (blue circle), $1.11\pm0.02$ (red square), $1.75\pm0.04$ (grey triangle).
For each temperature all points lie in the same Jarzynski manifold (which in this case is a line).
}
\end{figure*}


\end{document}